\documentclass[a4paper,fleqn,usenatbib]{mnras}
\usepackage{newtxtext,newtxmath}
\usepackage[T1]{fontenc}
\usepackage{ae,aecompl}

\usepackage{graphicx}%, epstopdf}	% Including figure files
\usepackage{amsmath}	% Advanced maths commands
\usepackage{amssymb}	% Extra maths symbols

\def\msun{\mbox{$M_\odot$}}

\def\mstar{\mbox{$M_{*}$}}

\def\re{\mbox{$r_{1/2}$}}
\def\Re{\mbox{$R_{e}$}}
\def\r3{\mbox{$r_{3}$}}
\def\vlos{\mbox{$\sigma_{\rm los}$}}

\title[Mass Estimators vs. Cosmological Simulations]{Dwarf Galaxy Mass Estimators vs. Cosmological Simulations}

\author[Gonzalez-Samaniego, A., et al.]{Alejandro Gonzalez-Samaniego,$^{1}$\thanks{E-mail: alejandro.g@uci.edu}
James S. Bullock,$^{1}$
Michael Boylan-Kolchin,$^{2}$ 
\newauthor
Alex Fitts,$^{2}$
Oliver D. Elbert,$^{1}$
Philip F. Hopkins$^{3}$, Du\v{s}an Kere\v{s}$^{4}$ and  
\newauthor
Claude-Andr\'{e} Faucher-Gigu\`{e}re$^{5}$
\\
$^{1}$Department of Physics and Astronomy, University of California, Irvine, CA 92697, USA\\
$^{2}$Department of Astronomy, The University of Texas at Austin, 2515 Speedway, Stop C1400, Austin, Texas 78712-1205, USA\\
$^{3}$TAPIR, California Institute of Technology, Pasadena, CA, USA\\
$^{4}$Department of Physics, Center for Astrophysics and Space Sciences, University of California, San Diego, La Jolla, CA, USA\\
$^{5}$Department of Physics and Astronomy and CIERA, Northwestern University, Evanston, IL, USA\\
}

\date{Accepted XXX. Received YYY; in original form ZZZ}

\pubyear{2016}

\begin{document}
\label{firstpage}
\pagerange{\pageref{firstpage}--\pageref{lastpage}}
\maketitle

% Abstract of the paper
\begin{abstract}
We use a suite of high-resolution cosmological dwarf galaxy simulations to test the accuracy of commonly-used mass estimators from \citet[][]{Walker+2009} and \citet[][]{Wolf+2010}, both of which depend on the observed line-of-sight velocity dispersion and the 2D half-light radius of the galaxy, $\Re$.  The simulations are part of the the Feedback in Realistic Environments (FIRE) project and include twelve systems with stellar masses spanning  $10^{5} - 10^{7} M_{\odot}$  that have structural and kinematic properties similar to those of observed dispersion-supported dwarfs.  
Both estimators are found to be quite accurate:  $M_{\rm Wolf}/M_{\rm true} = 0.98^{+0.19}_{-0.12}$ and $M_{\rm Walker}/M_{\rm true} =1.07^{+0.21}_{-0.15}$, with errors reflecting the $68\%$ range over all simulations.  The excellent performance of these estimators is remarkable given that they each assume spherical symmetry, a supposition that is broken in our simulated galaxies.  Though our dwarfs have negligible rotation support, their 3D stellar distributions are flattened, with short-to-long axis ratios $ c/a \simeq 0.4-0.7$.  The accuracy of the estimators shows no trend with asphericity.    Our simulated galaxies have sphericalized stellar profiles in 3D that follow a nearly universal form, one that transitions from a core at small radius to a steep fall-off $\propto r^{-4.2}$ at large $r$; they are well fit by S\'{e}rsic profiles in projection.
We find that the most important empirical quantity affecting mass estimator accuracy is \Re .   Determining \Re\ by an analytic fit to the surface density profile produces a better estimated mass than if the half-light radius is determined via direct summation.   
%We find no preference for circular vs. elliptical binning of the light profile in determining \Re\ for use in these mass estimators.
\end{abstract}

%%%%%%%%%%%%%%%%% BODY OF PAPER %%%%%%%%%%%%%%%%%%

%=========================================================
\section{Introduction}
\label{intro}
%=========================================================

Dynamical measurements of galaxy masses are among the most important empirical quantities for testing theories of galaxy formation and cosmology (see \citealt{Courteau+2014} for a review).   In rotationally-supported systems, for example, velocity field measurements can be used to trace the total mass enclosed as a function of radius, and this provides an important test of $\Lambda$CDM predictions for dark matter halo density profiles shapes \citep[e.g.,][]{flores1994,deBlok2001,oh2010,Iorio+2016}.  
 But for dispersion-supported galaxies, radial mass profiles are harder to extract from line-of-sight velocities, owing significantly to degeneracies associated with the unknown velocity dispersion anisotropy \citep[e.g.,][]{Lokas+2002,BT08}.  The inability to extract mass-density profiles in a straightforward manner from dispersion-supported galaxies is particularly unfortunate in the case of small (\mstar $\lesssim 10^7$ \msun) dwarf spheroidal/irregular galaxies    as these systems are the primary focus of the Missing-Satellites problem
\citep{klypin+1999,Moore+1999,SG07,Brooks2013} and the Too-Big-to-Fail problem \citep{BK+2011,BK+2012,GK+2013, tollerud+2014}.

The mass-anisotropy degeneracy can in principle be broken by higher-order velocity moments \citep[e.g.][]{Lokas+2005,Cappellari+2008} and/or proper motion data \citep{Strigari+2007,Read+2017}.  %Efforts of this kind are important, though they rely on an assumption
%that the Jeans equation holds, which may not be the case for galaxies that have had significant recent feedback episodes \citep{badry+2016}.
A common and easy-to-implement practice is to use the observed line-of-sight velocity dispersion to infer the mass within a  single characteristic radius, close to the galaxy's half-light radius, where mass measurements are most robust to degeneracies associated with velocity dispersion anisotropy.  This approach was specifically advocated by  \citet[][Wa09]{Walker+2009} and \citet[][Wo10]{Wolf+2010}, who presented simple estimators designed for dwarf galaxies in particular.  These estimators have become the go-to empirical measures of mass used in confronting the Too-Big-to-Fail problem \citep[][]{Fattahi+2016,Wetzel+2016} and in reporting mass-to-light ratios of newly discovered dwarfs to distinguish baryon-dominated star clusters from dark-matter dominated ultrafaint galaxies \citep[][]{Willman+2012,Voggel+2016,Caldwell2016}.

In this paper, we will test these common mass estimators against fully self-consistent cosmological simulations of \mstar $\approx 10^5 - 10^7$ \msun~ dwarf galaxies. Our simulations are part of a FIRE-2 simulation suite of the Feedback in Realistic Environments (FIRE) project \citep{hopkins+2014,fitts+2016,hopkins+2017}\footnote{\url{http://fire.northwestern.edu}}.  We note that \citet{badry+2016} have used FIRE simulations to test the underlying reliability of the Jeans equation in galaxies experiencing feedback and concluded that while larger dwarfs (\mstar $> 10^7$ \msun) can experience feedback-driven potential fluctuations that induce $\sim 20\%$ errors in full Jeans modeled masses, the single small galaxy they considered (\mstar $\sim 2.5\times10^{6}$ \msun) had only a $5\%$ offset in derived mass from Jeans modeling owing to the rather modest energy injection it had received from supernovae.  This result provides hope that mass estimators we aim to test will prove reliable when compared against our larger suite of low-mass dwarfs.

Both the Walker and Wolf estimators were derived using idealized approximations, including the assumption that galaxies are spherically symmetric and non-rotating.  While small dwarfs do tend to have little stellar rotation \citep{Wheeler+2017} , they can be quite aspherical \citep{mcconnachie2012}, and this in itself motivates an exploration of the estimators' reliability.  Past tests of this kind have relied on analytic approaches \citep{sanders+2016} and mock observations of galaxy-formation simulations  \citep{Campbell+2016}.  \citet{sanders+2016} showed that spherically-symmetric mass estimators can be applied to flattened stellar distributions, if \Re\ is replaced by the so called ``circularized' analogue \Re$\sqrt{1-\epsilon}$, where $\epsilon$ is the ellipticity of the projected stellar distribution. Using the {\it APOSTLE} simulations that naturally included non-spherical galaxies, \citet{Campbell+2016} found that both Wa09 and Wo10 provide accurate dynamical masses for their simulated galaxies with $\sim 25\%$ (unbiased) offsets at one-sigma.  
Our approach is similar, though complementary, to that of \citet{Campbell+2016}.  While their simulations included a very large number of galaxies over a range of stellar masses (\mstar = $10^{6.5} - 10^{10}$ \msun) that are generally larger than most of the Milky Way satellites, ours includes $12$ systems with masses that are better matched to the range of classical Local Group dwarfs (\mstar = $10^{5.6} - 10^{7.2}$ \msun). %What we lack in numbers we make up for in resolution: 
Our simulations also have $\sim 15 \times$ higher mass resolution and $\sim 3 \times$ better force resolution\footnote{In GIZMO, we use the same definition of force softening $\epsilon$ and spatial resolution, corresponding to the inter-particle separation $h_{i}$; this is related to the Plummer equivalent softening  ($\epsilon_{{\rm plummer}}$) by $\epsilon_{{\rm plummer}} \approx (2/3)\epsilon$.} ($\epsilon_{{\rm DM}} = 35$ pc) than the best-resolved systems in \citet{Campbell+2016}.
High resolution is crucial for this kind of analysis, as the galaxies of concern are small (\Re\ $\sim 500$~pc) and the mass estimators are proportional to galaxy size. Thus our analysis extends the mass-estimator test to the mass-scale of most relevance to the Missing Satellites and Too-Big-to-Fail problems and further allows us to perform a careful system-by-system analysis of each well-resolved galaxy.  Finally, by testing these mass estimators with an entirely different simulation code and different modeling of galaxy formation physics, we provide an all-important cross check to ensure robustness to underlying model implementations and assumptions.

This article is organized as follows.  In Section \ref{method} we present our suite of simulations and discuss the methods to test mass estimators. In Section \ref{results}  we present our main results, which include a summary of the important properties of our simulated galaxies in \ref{properties} and the mass estimator performance on each galaxy in \ref{inSims}.   Section \ref{discuss} summarizes our findings, compares them to previous work, and discusses future directions.

%=========================================================
\section{Methods}
\label{method}
%=========================================================

%=========================================================
\subsection{Mass estimators and assumptions}
\label{assum}
%=========================================================

The dynamical mass estimators we will test were derived from the spherically-symmetric Jeans equation:
\begin{equation}
\frac{1}{n_{*}} \frac{d}{dr}\left(n_{*} {\sigma^{2}_{r}} \right)+ 2\frac{\beta{\sigma^{2}_{r}}}{r} = -\frac{GM(r)}{r^{2}},
\label{jeans}
\end{equation}
where $M(r)$ is the total mass within radius $r$.   The quantities $n_{*}$, $\sigma_{r}$, and $\beta \equiv 1 - \sigma_t^2/\sigma_r^2$ represent the stellar number density, the radial velocity dispersion, and the velocity dispersion anisotropy, respectively. The case of $\beta =0$ implies that the tangential velocity dispersion $\sigma_t$ ($= \sigma_\theta = \sigma_\phi$) is equal to the radial velocity dispersion.   It is revealing to rewrite Equation \ref{jeans} by solving for the mass profile and defining $\gamma_*  \equiv - {\rm d}\ln n_*/{\rm d}\ln r$ and $\gamma_\sigma \equiv - {\rm d}\ln \sigma_r^2/{\rm d}\ln r$ to give
\begin{equation}
M(r) = \frac{r \sigma^{2}_{r}}{G} \left(\gamma_* + \gamma_\sigma - 2\beta \right) \, .
\label{jeansM}
\end{equation}
We see that even at fixed radial velocity dispersion and radius, the associated mass is very sensitive to the local slopes of the tracer profile,  the tracer's velocity dispersion profile, and the velocity dispersion anisotropy.  Each of these quantities is difficult to measure in a galaxy observed in projection.

\begin{figure*}
%\begin{tabular}{ll}
\hspace{.2cm}\includegraphics[width=8.2cm,height=6.45cm]{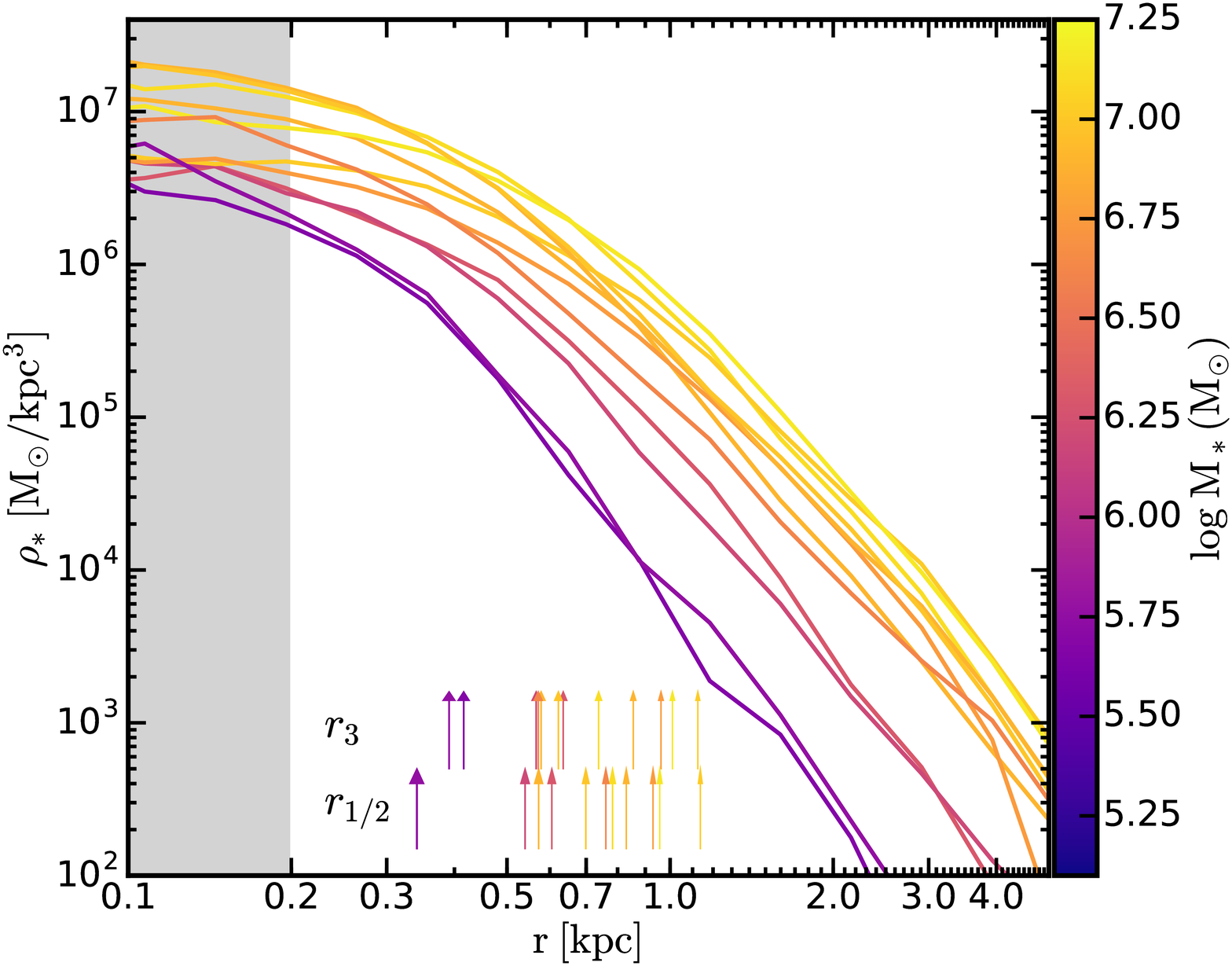} 
\hspace{.2cm}\includegraphics[width=8.2cm,height=6.4cm]{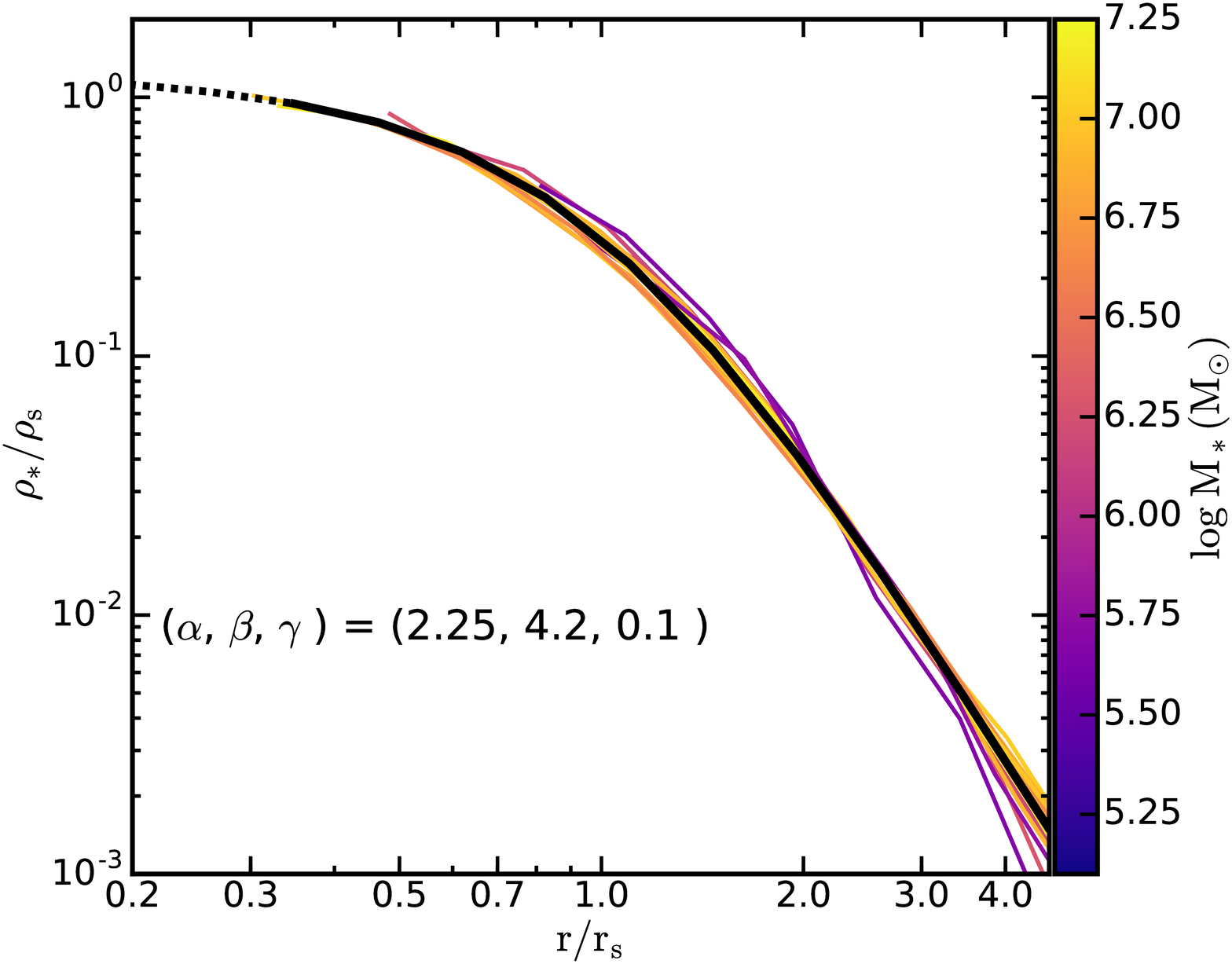} 
%\end{tabular}
  \caption{Left: Stellar density profiles for the dwarf galaxies in our suite of simulations. The color code denotes the galaxy's stellar mass.  The arrows mark the radius where the log-slope of the profile equals $-3$ (\r3) and the 3D half-mass radius ($r_{1/2}$), as indicated.  Note that the Wo10 estimator, as usually applied, assumes that $r_{1/2} \approx \r3$.  This is a good approximation for most of our simulations but can break down in some cases, with implications for mass estimator accuracy (see Section~\ref{inSims}). The grey shaded area marks the region where the innermost profiles may be affected by numerical relaxation %likely are not dynamically resolved 
 \citep[see][]{fitts+2016}. Right: the same profiles normalized to a single-shape
 $\alpha\beta\gamma$-profile (Eq. \ref{zhao}) with $(\alpha, \beta, \gamma) = (2.25, 4.2, 0.1)$.  The densities have been normalized to by the constant $\rho_s$ and the radii have been normalized to the characteristic scale radius $r_s$, where $r_s  \approx 1.2 \, \r3$ for the the best-fit shape parameters (Eq. \ref{sprofile}). Each galaxy's stellar mass profile has been truncated within $r \simeq 200$ pc, reflecting a conservative assessment of convergence, though the fit remains good within these radii as well.
 }
\label{densities}
\end{figure*}

Both Wa09 and Wo10 used Equation \ref{jeans} as a starting point and derived mass estimators of the form $M(<r_{\star}) = \alpha r_{\star} \, \sigma^{2}_{\rm los} / G$, where
$r_{\star}$ is a characteristic stellar radius within which the mass is measured, $\sigma_{\rm los}$ is the tracer-weighted {\em average} line-of-sight velocity dispersion, and $\alpha$ is a constant of order unity.

 Wa09 were guided by MCMC-enabled spherical Jeans modeling of their observed galaxies, which showed that masses within a radius $r_{\star} = \Re$ (the projected half-light radius) were generally well constrained in posterior likelihoods after marginalizing over many other parameters.  They therefore worked out analytically the mass within $\Re$ for spherical system simplified by three assumptions: i) that $n_*$ is given by a Plummer profile, ii) that $\beta=0$, and iii)  that $\gamma_\sigma =0$.  They obtained 
\begin{equation}
M_{\rm Walker} (< \Re) = \frac{5 \Re \sigma^{2}_{\rm los}}{2\, G}, 
\label{eqEs1}
\end{equation}
and showed that this estimator did well in comparison to their full Jeans/MCMC analysis.

Wo10 set out to find an ideal radius to use for $r_{\star}$.  Assuming $\gamma_\sigma = 0$, they were able to derive analytically that the mass within a radius
$r_{\star} = \r3$ is independent of $\beta$ for an observed $\sigma_{\rm los}$.  Here, $\r3$ is the radius where $\gamma_* = 3$ (the radius where the the log-slope of the tracer profile is equal to $-3$).  The equation they derived is
\begin{equation}
M_{\rm Wolf}^{\rm ideal} (< \r3) = \frac{3 \r3 \sigma^{2}_{\rm los}}{G}.
\label{eqEs2}
\end{equation}
Wo10 also showed that for a wide range of commonly-adopted stellar profiles that $\r3 \simeq r_{1/2}$, where $r_{1/2}$ is the 3D half light radius.  The Wolf formula is then often quoted as an estimator for $M(<r_{1/2})$.  Another good approximation that allows one to determine $r_{1/2}$ from the projected half-light radius $\Re$ is $r_{1/2} \simeq 4/3 \Re$ (see Wo10).  Since $\Re$ can be measured on the sky, this is the most common form of the Wolf estimator used in the literature and we will adopt it as our fiducial comparison case:
\begin{equation}
M_{\rm Wolf}(< 4/3 \Re) = \frac{4 \Re \sigma^{2}_{\rm los}}{G}.
\label{eqEs3}
\end{equation}

In what follows we will test Equations \ref{eqEs1} and \ref{eqEs3} using mock observations of our simulations.

%=========================================================
\subsection{Simulations}
\label{sims} % used for referring to this section from elsewhere
%=========================================================

%This work relies on a suite of twelve $\Lambda$CDM  zoom-in hydrodynamical simulations of dwarf galaxies introduced by \citet{fitts+2016}.  The simulations were run using
%the GIZMO code \citep{hopkins+2015} with FIRE-2 physics \citep{hopkins+2017}, which updates from the original Feedback In Realistic Environments (FIRE) project  \citep{hopkins+2014}.  The FIRE-2 implementation has been shown to reproduce observables from real galaxies with success 

This work relies on a suite of twelve $\Lambda$CDM zoom-in hydrodynamical simulations of dwarf galaxies \citep{fitts+2016} that were run using the GIZMO code\footnote{\url{http://www.tapir.caltech.edu/~phopkins/Site/GIZMO.html}} \citep{hopkins+2015}. The simulations are part of the Feedback In Realistic Environments (FIRE) project \citep{hopkins+2014} and adopt the FIRE-2 model for galaxy formation physics \citep{hopkins+2017}. FIRE-2 implements the same main star formation and stellar feedback physics models as the original FIRE simulations, but with several numerical improvements, including a more accurate hydrodynamics solver (the ``meshless finite mass'' method). FIRE galaxies successfully reproduce several observables from real galaxies \citep[see, e.g.,][]{onorbe+2015,Chan+2015,Wheeler+2017,muratov+2015,FG+2015,FG+2016,ma+2016,badry+2016b}.  
We refer the reader to \citet{fitts+2016} for  a complete description of the simulations used in this work. Here we only briefly summarize the main characteristics.

  The FIRE-2 simulations include cooling and heating rates tabulated from CLOUDY \citep{ferland+2013}. Star formation proceeds only in the densest regions ($\ge 1000 \, {\rm cm}^{-3}$), mimicking the densities of molecular clouds.  Each star particle is spawned as a single stellar population with a \citet{kroupa+2001} initial mass function. The simulations model momentum and energy input from stellar feedback, including photo-ionization and photo-electric heating, radiation pressure, stellar winds, and supernovae (Types Ia and II).  

Our simulated galaxies each form within a dark matter halo of virial mass $10^{10} M_{\odot}$ at $z=0$ with their properties summarized in Table 1 of  \citet{fitts+2016}.  Specifically, we are studying simulations m10b to m10m.  The initial conditions for these runs were chosen to span a variety of formation histories and associated halo concentrations and this results in a range of stellar masses in the final timestep:  $M_{*} = 5 \times10^{5}$ to  $1.5 \times 10^{7} \msun$.   The simulations use gas particle masses of 500 $\msun$ (with star particles having somewhat lower masses) and dark matter particles of mass 2500 \msun.  The force resolution is $\epsilon_{DM}=35$ pc (corresponding to a ``Plummer-equivalent'' softening of 23 pc). The minimum gravitational and hydrodynamic resolution reached by the gas is $\epsilon_{g}=h_{g}=2$ pc, which corresponds to the inter-particle separation. Stars have a fixed force resolution at $\epsilon_{\star}=2$ pc. This level of mass and force resolution allows us to resolve the internal structure of simulated galaxies with thousands of star particles  -- and thus with tracers comparable in number to the largest kinematic stellar samples in local dwarf galaxies.   As discussed 
in \citet{fitts+2016}, these simulated dwarf galaxies have properties that broadly reproduce many observed properties of real galaxies of the same stellar mass and are highly dark-matter dominated.  
Their dark matter density profiles vary in shape from (feedback-driven) cores in systems with \mstar $\gtrsim 3\times 10^{6} \msun$ to cuspy profiles in galaxies that formed fewer stars than this threshold.  This diversity in the underlying dark matter profile shape in our simulated systems adds to the generality of the tests of mass estimators.

\begin{figure} 
\hspace{-0.35cm}\includegraphics[scale=0.262]{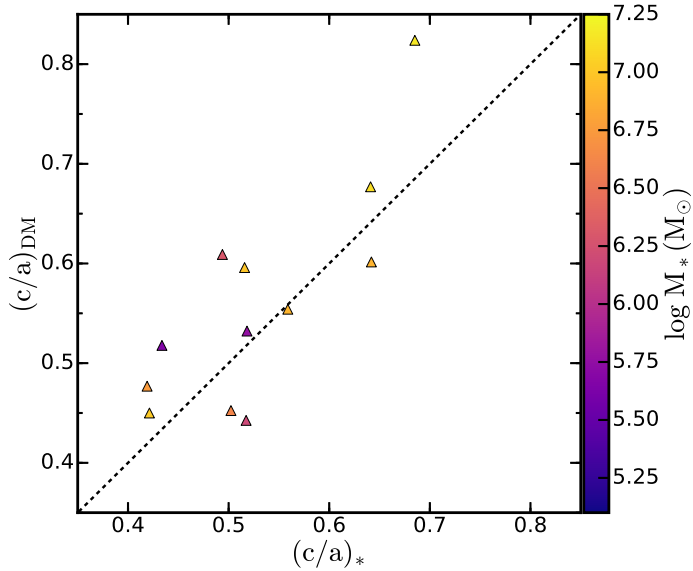} 
  \caption{Points show the 3D short-to-long axis ratios of dark matter $(c/a)_{\rm DM}$ and stars $(c/a)_{*}$ in our simulated galaxies, measured iteratively within central volumes equal to that of a sphere of radius $r_{1/2}$.  Colors correspond to stellar mass as indicated in the color bar.   While most of the simulated galaxies sit close to the one-to-one line (dashed), the shapes of stars and dark matter are not identical. } 
\label{fig:shape}
\end{figure}

\begin{figure}
    \hspace{-0.2cm} \includegraphics[scale=0.25]{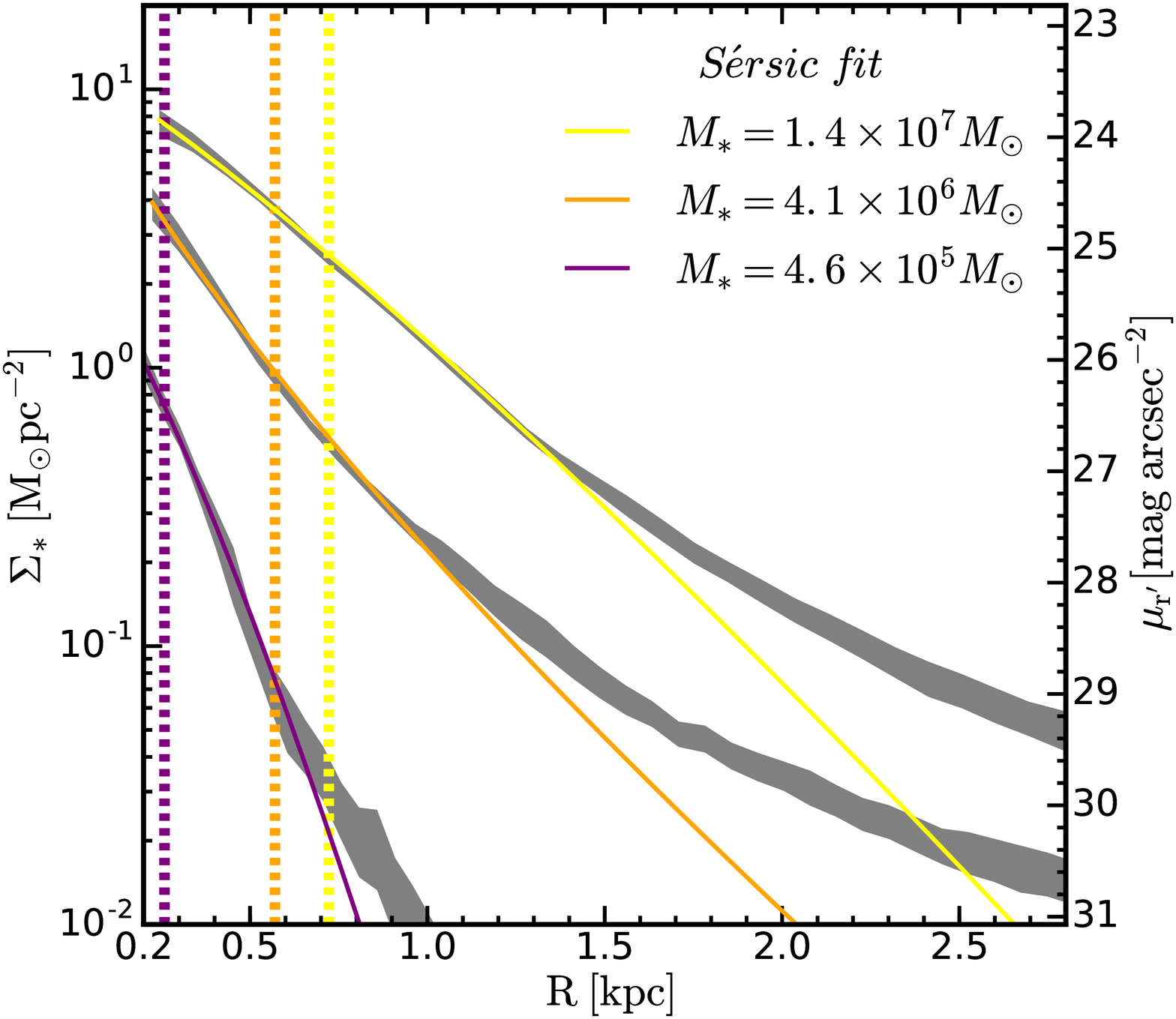}
    \caption{Surface density profiles for three of the simulated dwarfs plotted as a function of projected radius $R$. These three galaxies span our full range of stellar densities in the simulation suite. The grey bands show the $1\,\sigma$ range over the resultant profiles from 1000 random projections sampling the unit sphere for each galaxy and the dispersion is owing to asphericity.
The solid lines (colored according to the stellar mass of the galaxy) represent the S$\'{e}$rsic fit for circular shells to an average projection in each galaxy; \Re\ inferred by fitting each projection is represented by the vertical lines with the width corresponding to the $1\,\sigma$ range. Importantly, while the preferred S\'{e}rsic index fits do vary from projection to projection, the resultant  \Re\ values (which matter most for the mass estimators) are quite stable to viewing angle.  Note that the brighter galaxies display an upward break at large radii -- such an extended `stellar halo' type feature could in principle be observable around dwarf galaxies at very low surface brightness.  We assumed $M_{*}/L_{r^{\prime}} =1$ to convert $\Sigma_{*}$ to surface brightness in the $r^{\prime}$-band to make this figure.}%The horizontal black dotted line represent the equivalent surface brightness $\mu_{r^{'}} = 26.2$ mag arcsec$^{-2}$.}
    \label{surface}
\end{figure}

%=========================================================
\section{Results}
\label{results}
%=========================================================

\subsection{Properties of simulated galaxies}
\label{properties}

As discussed in \S \ref{assum}, the mass estimators we will be testing were derived under the simplifying assumptions of spherical symmetry and a constant stellar velocity dispersion profile. The Jeans equation is also sensitive to the shape of the underlying tracer density profile, and the Wo10 estimator in particular relies on the assumption that $\r3 \simeq r_{1/2} \simeq 4/3 \Re$.  Each of these assumptions is broken to some degree by our simulated galaxies, and in this section we briefly present results on the stellar density distributions (both in 3D and in projection), the velocity dispersion profiles, and the ellipticity of our simulated galaxies before moving on to test the mass estimators applied against them.  We note that our simulated galaxies display negligible stellar rotation, with very similar dispersion-supported properties to the simulated FIRE dwarfs discussed in \citet{Wheeler+2017}.

The left panel of Figure~\ref{densities} presents the spherically-averaged stellar density profiles of each of our simulated galaxies.    The color code corresponds to the total stellar mass of each galaxy, as shown by the color bar along the right.  Arrows mark the radius where the local log-slope of the density profile equals -3 (\r3), and the 3D-half mass radius $r_{1/2}$, respectively.   The size range spans $\sim 400 - 1100$ pc and is consistent with the sizes of real dwarf galaxies in this mass range, as discussed in \citet{fitts+2016}.  Recall that the Wo10 estimator assumes that $\r3 \approx r_{1/2}$ and this is, in fact, fairly accurate for almost all of the simulations.  For nine of the twelve galaxies, the difference is smaller than $10\%$.   The grey band marks the region where predictions are potentially unreliable due to numerical relaxation \citep[][]{fitts+2016}.

\begin{figure} 
\hspace{-0.5cm}\includegraphics[scale=.28]{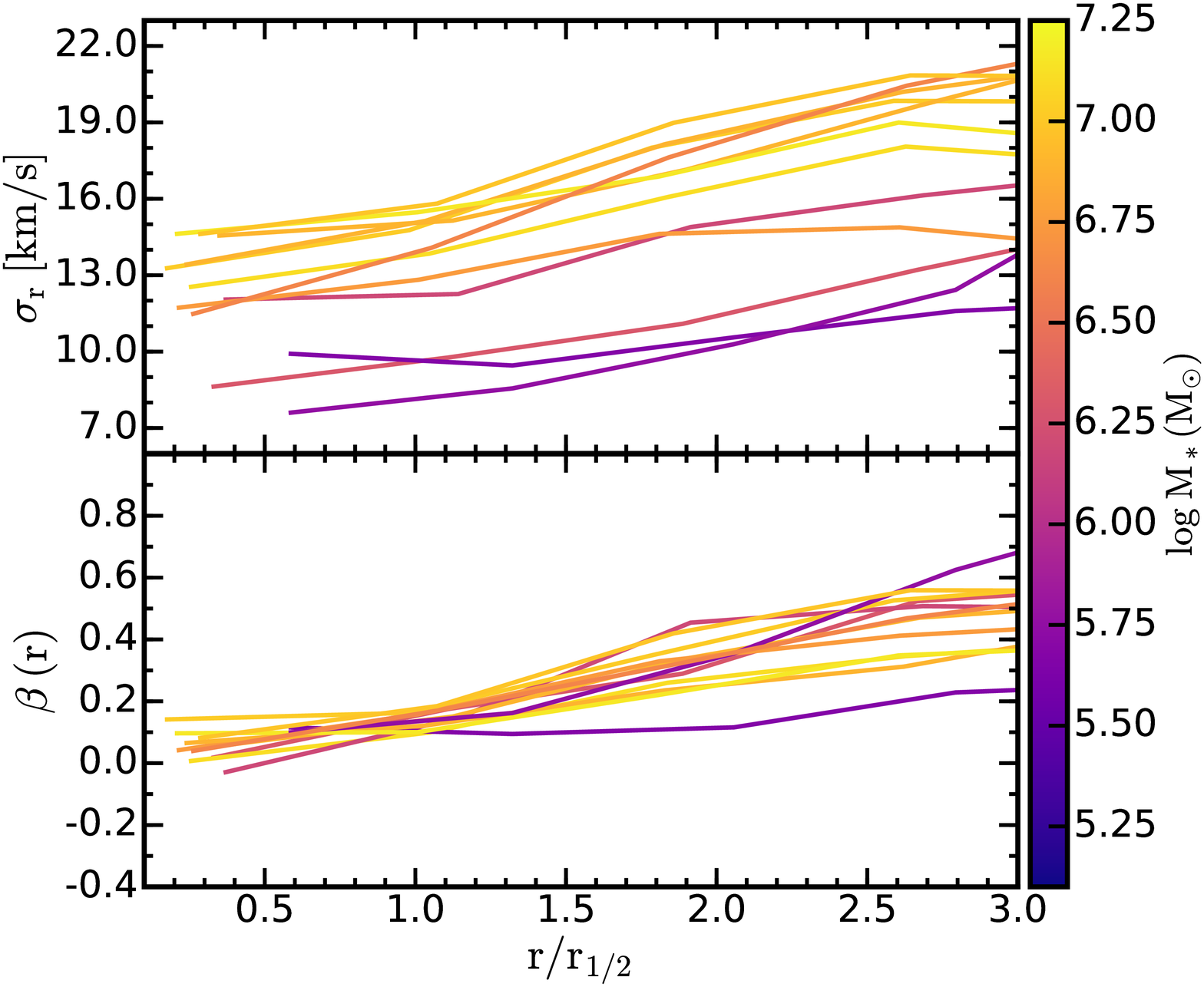} 
  \caption{Top: Stellar radial velocity dispersion profile for each of our simulated dwarf galaxies. The horizontal axis has been normalized to the 3D half mass radius \re . We see that our $\sigma_r$ profiles rise gradually towards larger radius.  Bottom: The stellar velocity dispersion anisotropy profiles $\beta(r)$ for the same simulations.   Most systems are fairly isotropic towards the center ($\beta \sim 0$) with increasing radially-biased velocity dispersions ($\beta > 0$) towards large radii.  }
\label{sigma_r_beta}
\end{figure}

\begin{figure} 
\hspace{0.0cm}\includegraphics[scale=.28]{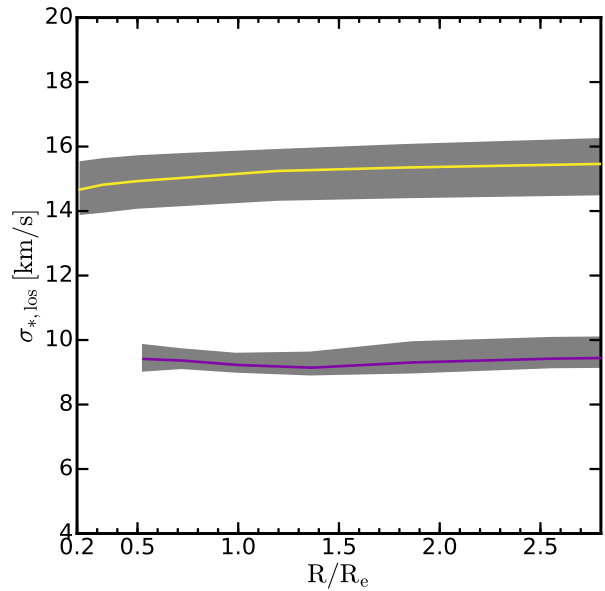} 
  \caption{Line-of-sight velocity dispersion profiles for two of our simulated dwarfs. The two span the full range of stellar masses in the simulation suite: $M_{*} = 5 \times10^{5}$ to $1.5 \times 10^{7} \msun$. The grey bands show the $1\,\sigma$ range over the resultant profiles from 1000 random projections covering the unit sphere for each galaxy, with the width of the band driven by the non-spherical nature of each galaxy's velocity ellipsoid.  The solid colored lines show the median over all projections.  The observed profiles are flat, similar to those of observed dwarf galaxies.}
\label{sigma_los}
\end{figure}

As Figure~\ref{densities} makes clear, the shapes of the $\rho_*(r)$ distributions in our simulated dwarf galaxies are quite similar.  We find that all of them are reasonably well approximated
by a single ``universal" shape.  Specifically, we have
fit an $\alpha\beta\gamma$-profile \citep{zhao+1996} of the form 
\begin{equation}
\rho(r) = \frac{\rho_{s}}{(r/r_{s})^{\gamma}\left[ 1 + (r/r_{s})^{\alpha}\right]^{(\beta-\gamma)/\alpha}}\, ,
\label{zhao}
\end{equation}
to each galaxy and find a good fit to the sample as a whole to be:
 \begin{equation}
\rho_*(r) = \frac{\rho_{s}}{(r/r_{s})^{0.1}\left[ 1 + (r/r_{s})^{2.25}\right]^{1.82}}.
\label{sprofile}
\end{equation}
This profile asymptotes to a core-like distribution at small radius ($\propto 1/r^\gamma$ with $\gamma = 0.1$) and a steep fall-off at large radius ($\propto 1/r^{\beta}$ with $\beta = 4.2$) and obeys  $\r3 \simeq \, 0.85 \, r_s$.   In the right panel of Figure~\ref{densities}, we plot the density profiles for each simulated system normalized in radius by $r_s$ and in density by $\rho_s$. The black line shows Eq. \ref{sprofile}, which goes dashed in the innermost radius of the galaxies, which may not be well resolved.  Though this parameterization does a good job of characterizing the general shape of our galaxies, it is worth noting that the fit is not perfect: the smallest galaxies, for example, prefer somewhat steeper outer profiles.

The density profiles shown in Figure~\ref{densities} enforce spherical symmetry, but in reality our simulated galaxies are non-spherical.  We have computed the shape tensors $\mathbf{S}$ \citep[as described in][]{Zemp+2011} for stars and dark matter within our galaxies and diagonalized  them to obtain their eigenvalues, which are proportional to the square-root of the semi-principal axes of the 3D-ellipsoid $a\geq b\geq c$. While keeping the volume of an initial sphere  of radius $r$ constant, we use an iterative method to construct ellipsoids until the axis ratio is stable to $1\%$.  The points in Figure~\ref{fig:shape} show short-to-long
axis ratios $c/a$  (where $c/a = 1$ is perfectly spherical) of dark matter vs. stars calculated in volumes defined by spheres of radius $r=r_{1/2}$ for each simulated galaxy.  Despite their lack of rotational support, our simulated galaxies are quite flattened ($c/a \approx 0.4-0.7$).  The stellar and dark matter shapes are not identical, but the major axes of stars and dark matter align within 10 degrees for 11 of the 12 galaxies.  The most misaligned system has a 20 degree offset between the major axes of dark matter and stars.

\begin{figure}
	\hspace{0.2cm}\includegraphics[scale=0.365]{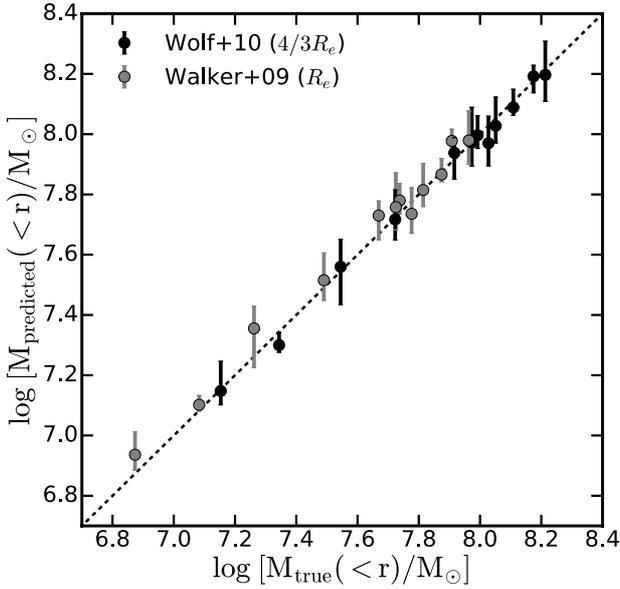}
    \caption{Estimated mass as a function of the true mass for each of our simulated dwarfs for the \citet{Wolf+2010} estimator (black circles) and for the \citet{Walker+2009} estimator (grey circles). The symbols represent the median values of 1000 line-of-sight measurements for each galaxy. The error bars are the 16th and 84th percentiles in the distributions.   Note that the Walker estimator measures mass within a slightly smaller radius than the Wolf estimator, thus the Walker-derived masses are shifted to the left and down relative to the Wolf-derived masses in each galaxy.
    }
    \label{one2one}
\end{figure}

\begin{figure}
\hspace{0.05cm}\includegraphics[scale=0.27]{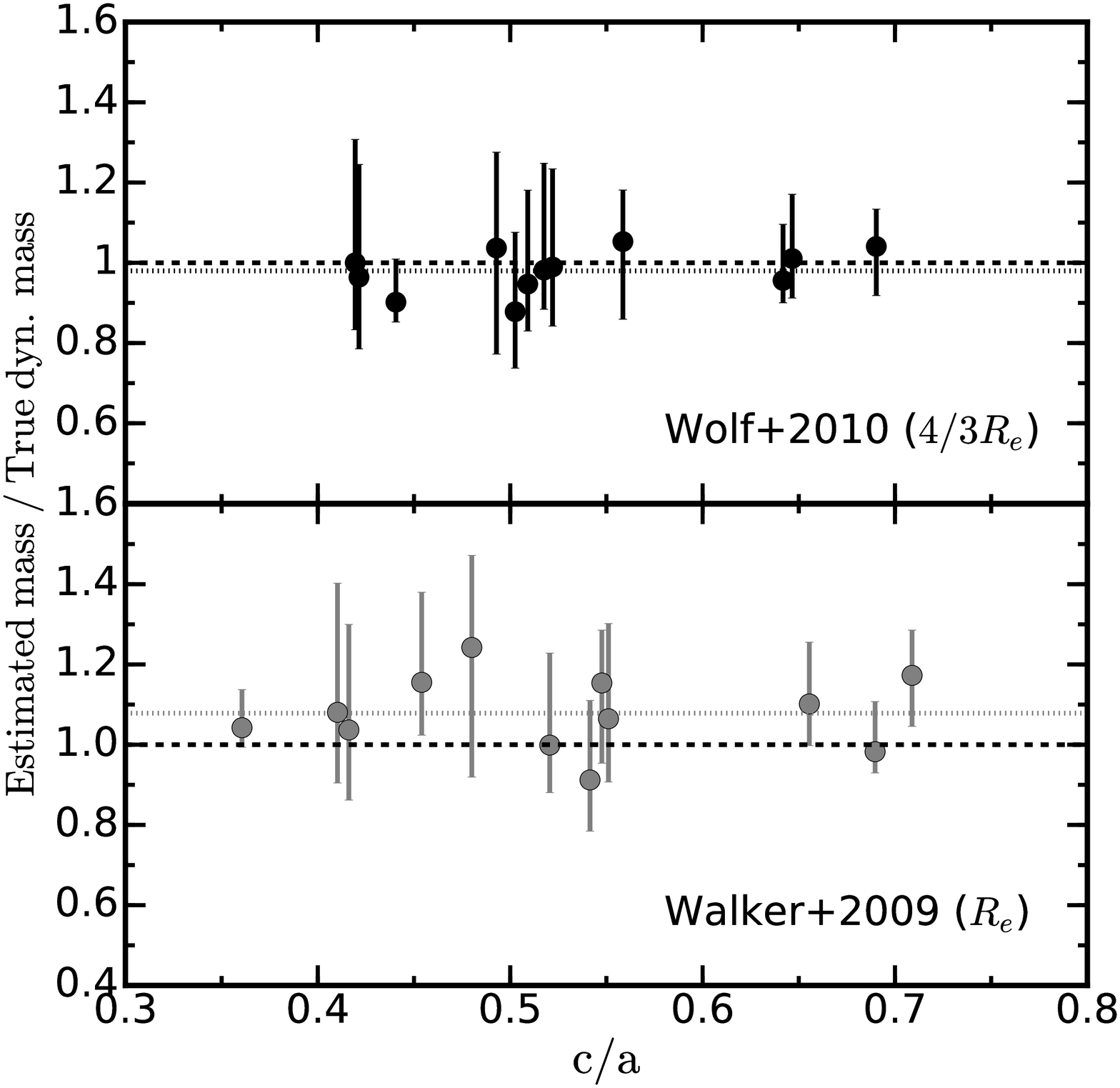}
    \caption{Ratio of estimated mass to true dynamical mass as a function of short-to-long axis ratio $c/a$ of stars in our simulated galaxies.  The top panel shows results for the  \citet{Wolf+2010} estimator (with $c/a$ and $M$ measured within $r_\star = 4/3 \Re$) and the bottom panel shows results for the \citet{Walker+2009} estimator (with $c/a$ and $M$ measured within $r_\star = \Re$).  As in Fig. \ref{one2one}, points reflect the median 
   over 1,000 random line-of-sight projections for each galaxy and the error bars show the $16-84$ percentile range over those projections.  The dash-dotted lines show the average of the medians for each galaxy in their respective panels. While median estimator accuracy shows no trend with $c/a$, in general the most aspherical systems do tend to produce more scatter about the median over multiple projections, as might be expected. Note that the points on the top and bottom axes do not necessarily have the same $c/a$ values because some of our galaxies have axis ratios that galaxy shapes that vary as a function of ellipsoidal radius.}
    \label{triaxial}
\end{figure}

While the 3D shape of the underlying stellar density distributions have important implications for full Jeans modeling, in practice it is the 2D (projected) profile that is observed and that must be used when inferring masses.  Specifically, the 2D half-light radius \Re\  is the only stellar distribution parameter that is explicitly included in the Walker and Wolf formulas (Eq. \ref{eqEs1} and \ref{eqEs3}).   Figure~\ref{surface} shows the surface density profiles for three of our simulated dwarfs plotted as a function of 2D (circularly-averaged) radius. The shaded bands show the $1\,\sigma$ range of density profiles observed over 1,000 random projections covering the unit sphere around each galaxy. The three colored lines show 
average S\'{e}rsic profile fits over all projections for each galaxy (color coded by galaxy \mstar).  We perform these fits within $R = 4 \,r_{1/2}$.  In general, our simulated surface density profiles are well-characterized by S\'{e}rsic fits with $0.7\lesssim n \lesssim 1.3$ in their inner regions ($r \lesssim 2-4$ \Re).  In this sense, our galaxies are realistic analogs for testing mass estimators.  Interestingly, while a single galaxy can yield a variable S\'{e}rsic index $n$ when viewed from different projections, the resultant 2D half-light \Re\ values {\em derived from the fits} are quite robust from projection-to-projection.  The color-matched vertical dotted lines show $\pm 1 \sigma$ range of \Re\ values for each galaxy over all projections.  This narrow range of inferred \Re\ for each galaxy contributes to the accuracy of the mass estimators (as we discuss below).

Note that the two most massive systems shown in Figure~\ref{surface} demonstrate an outer flattening in their projected light profiles. We see this behavior -- an excess above the S\'{e}rsic extrapolation at large radius --  in most of the more massive systems in our suite. Similar features have been observed in the outer regions of larger disk galaxies \citep[see e.g.,][]{herrmann+2016}, and it is possible that behavior of this kind exists in the outskirts of smaller dwarf galaxies such as the ones simulated here. However detecting such a flattening in a diffuse stellar component could be challenging.  With resolved stars, such a break might be hard to distinguish from foreground or background stars.  In diffuse light, the fairly low surface brightness of the expected upturn will make it difficult to detect. In Figure~\ref{surface} the equivalent surface brightness $\mu_{r^{'}}$ in the SDSS $r'$ band is shown in the right-hand axis. Observational searches for behavior of this kind in the outer regions of dwarf galaxies could enable an interesting constraint on simulations of the type presented here.

The top panel of Figure~\ref{sigma_r_beta} shows the radial velocity dispersion profiles of each simulated dwarf galaxy.  The bottom panel shows the velocity anisotropy profiles for the same systems.  The radii have been normalized by $\re$ in each case.  The radial velocity dispersion profiles rise slowly towards larger radius.  The velocity dispersion anisotropy also rises from nearly isotropic $\beta \sim 0$ at small $r$ to a radially biased velocity dispersion at large $r$.  The fact that $\sigma_r$ is not constant for our galaxies provides an interesting test for both the Wa09 and Wo10 estimators, given that both assume $\gamma_\sigma \equiv$ - d$\ln \sigma_r^2$/d$\ln r = 0$.  Our galaxies, on the other hand, have $\gamma_\sigma \sim - 0.5$ near $ r \sim \re$.  

Figure~\ref{sigma_los} shows that while the $\sigma_r$ tends to rise with radius, our galaxies produce flat line-of-sight velocity dispersion profiles when observed in projection, similar to those seen in real dwarfs (see, e.g., Wa09).   In a similar approach as used in Figure~\ref{surface}, Figure~\ref{sigma_los} presents line-of-sight velocity dispersion profiles of two simulated dwarfs plotted as a function of 2D (circularly-averaged) radius (normalized by $\Re$). The shaded bands show the $1\,\sigma$ range of velocity dispersion profiles observed over 1,000 random projections covering the unit sphere around each galaxy. The two colored lines show 
the median over all projections for each galaxy (color coded by galaxy \mstar).

%\begin{figure}
%\hspace{-0.5cm}	\includegraphics[scale=0.25]{figures/triaxial_shape_sersicCyl_A.eps}
%    \caption{Ratio of estimated mass to true dynamical mass as a function of triaxiality of stars for our simulated galaxies.  The top panel shows results for the  \citet{Wolf+2010} estimator and the bottom panel shows results for the \citet{Walker+2009} estimator.  As in Fig. \ref{one2one}, points reflect the median over 1,000 random line-of-sight projections for each galaxy and the error bars show the $16-84$ percentile range over those projections.  The dotted lines show the average of the medians for each galaxy in their respective panels. {\bf[JB: Looks like the x-axis range / ticks are somehow wrong.  Maybe shift the "Wolf+" and "Walker+" labels to the bottom of each panel so that they don't overlap with error bars.]} }
%    \label{triaxial}
%\end{figure}

%=========================================================
\subsection{Testing the Walker and Wolf Mass Estimators}
\label{inSims}
%=========================================================

\begin{figure}
\hspace{-0.5cm}	\includegraphics[scale=0.36]{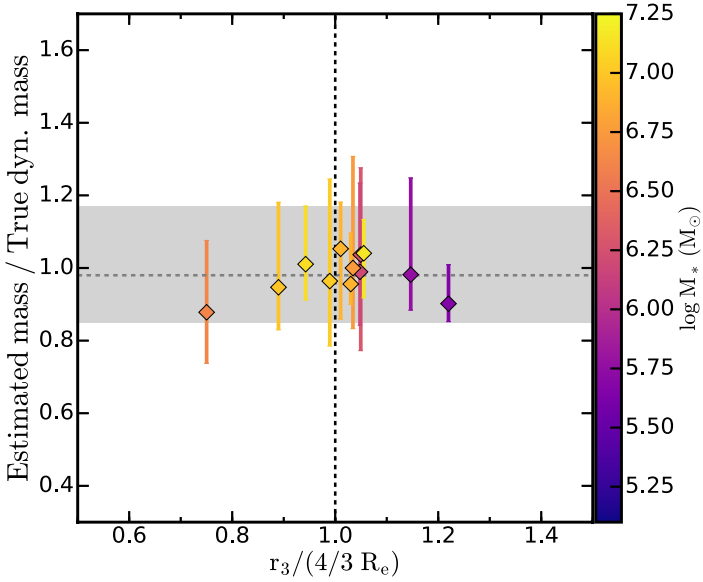}
    \caption{The ratio of Wolf mass to true dynamical mass  as a function of the ratio between the radius where the 3D density profile slope equals -3 ($r_{3}$) and 4/3 \Re .  We use the median $\Re$ over all projections (though the variance in $\Re$ is small).  As in Figure \ref{triaxial} the symbols represent the median values obtained from 1000 random projections over each simulated dwarf. The horizontal grey dashed line represent the median of the stacking distribution (see top panel in Figure \ref{stacking}) while the grey shaded region represent the 1-$\sigma$ uncertainty. Wo10 estimator assume that $r_{3} \approx 4/3 \Re$; this assumption fails at the $\sim 20\%$ level in the two galaxies where the estimator performs most poorly in the median ($\sim 10\%$ too low).}
    \label{r3re}
\end{figure}

\begin{figure}
	\hspace{-0.1cm}\includegraphics[scale=0.26]{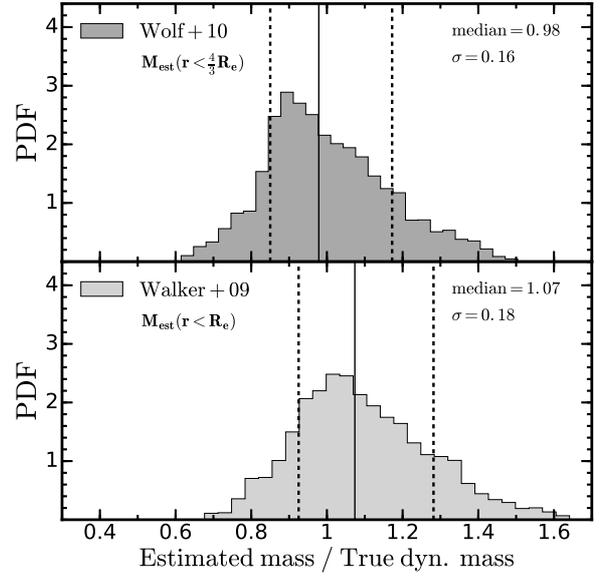}
    \caption{
    Probability distribution function of the ratio between the estimated mass and the true dynamical mass measured in the simulated galaxies. These distributions were obtained by stacking the 1000 random projections of each dwarf galaxy in our suite of simulations. Top panel: results obtained by using \citet{Wolf+2010} estimator, where the masses are measured up to $4/3$\Re. Bottom panel: a similar distribution plot is showed but now for \citet{Walker+2009} estimator, in this case the masses are measured up to \Re. In both cases \Re\ is measured by performing a S\'{e}rsic fit to each one of the 2D projections. The vertical dotted lines represent the 16th and 84th percentiles in the distributions.  The solid vertical lines are the medians.  The medians and $1 \sigma$ variance are shown in the legends of each panel.}
    \label{stacking}
\end{figure}

We perform mock observations of each of our dwarf galaxy simulations viewed along 1,000 randomly-sampled line-of-sight projections. For each projection, we compute the stellar-mass-weighted velocity dispersion, \vlos , and measure an \Re\ value using S\'{e}rsic fits to the circularly-binned stellar surface density profiles.   Figure~\ref{one2one} shows the resultant predicted mass obtained using both the Wo10 formula (black) and the Wa09 formula (grey) versus the true dynamical mass measured within the corresponding radii, $4/3\Re$ and \Re , respectively. Each symbol represent the median value over all projections.  There are two symbols for each galaxy, with the Wo10 mass always the larger of the two owing to the fact that it measures the mass within a larger radius.  The error bars  span the 16th and 84th percentiles of the distributions over all projections.   Note that each galaxy is plotted at a single `true' mass for either estimator.  This does not necessarily have to be the case, as the value of $\Re$ inferred from different projections could shift enough to force the true mass to shift accordingly.  However, as we discussed in \S  3.1, when we determine $\Re$ from fitting S\'{e}rsic profiles, the inferred distribution of $\Re$ values over all projections is extremely narrow, and this results in true total mass ranges (corresponding to ranges of $\Re$ value) that are narrower than the widths of the points plotted.

It is clear from Figure~\ref{one2one} that both estimators track the one-to-one line fairly well.   However, there are a few individual galaxies that tend to produce biased masses, with medians falling either above or below the line (slightly).   Figure \ref{triaxial} refines the comparison by showing the ratio of the estimated mass (Wolf, top; Walker, bottom) to true mass for each galaxy, again with points and error bars reflecting the median and one sigma range of the ratio, respectively. 
Each point is plotted against the $3D$ short-to-long axis ratio of the stars $c/a$ (as introduced for Fig. 3) measured within the radius $r_\star = 4/3 \Re$ (top) and $\Re$ (bottom) to reflect the radii of interest for the estimators.   
 The dotted line in each panel shows the average of the median data points for all galaxies.   We see no strong trend in the median points with $c/a$.

For the Wolf estimator (top), the median values are only $\sim 2\%$ low on average,  with the worst outliers biased low by $\sim 10\%$.  The one-sigma ranges are more important than the bias, with some galaxies having errors as large as $\pm 25 \%$ with respect to the true mass, though $\pm 15 \%$ is more typical.  The Walker estimator (bottom) does only slightly worse, with median values averaging about $\sim 10 \%$ high compared to the true mass.
The most extreme outliers are biased high in the median by $\sim 25 \%$ and the one-sigma range about the median is as large as $\pm 30\%$.  While these small discrepancies in the median do not correlate with $c/a$, many of the most aspherical galaxies are also the ones with the largest scatter. This is not surprising, since they will display the most pronounced variance in shape with viewing angle.   We have explored whether the offset in mass estimator correlates with
the {\em observed} projected shape on the sky and find no systematic trend (see Appendix ~\ref{shape}).  Unfortunately, based on our simulations, it is difficult to infer the expected variance based on the observed shape on the sky.

While asphericity in the stellar distribution does not seem to be the primary culprit in biasing estimators in the median, we do find that the mapping between 2D and 3D density -- specifically the relationship between \Re\ and $\r3$ -- is a source of bias.   Specifically, the Wo10 mass estimator assumes $r_{3} \approx 4/3 \Re$ because this is a good approximation for a variety of commonly-adopted analytic profiles (S\'{e}rsic, King, and Plummer).   However, this approximation does not have to hold in general.  In Figure~\ref{r3re} we plot the ratio of estimated Wolf mass to true dynamical mass against the $\r3/(4/3\Re)$ ratio derived for each system.   As before, each symbol is the median over 1,000 lines of sight for the twelve dwarfs. The error bars account for the 16th and 84th percentiles in the distribution.  We see that the two cases where the Wo10 estimator is biased ($\sim 10\%$ low in the median) are also the two cases where the $r_{3} \approx 4/3 \Re$ approximation  fails by more than 20 percent.  It therefore appears that the mapping between 2D stellar distribution and 3D distribution represents an underlying uncertainty in the mass measure, one that is difficult to overcome empirically.

In Figure~\ref{stacking} we show the probability distribution functions (PDF) obtained by stacking the ratios between the estimated mass and the true dynamical mass over all projections in all twelve simulated dwarf galaxies, with the Wolf estimator plotted in the upper panel and the Walker estimator plotted in the lower panel.  The solid vertical lines show the median of the distributions and the dotted lines show the 16th-84th percentiles:   $M_{\rm Wolf}/M_{\rm true} = 0.98^{+0.19}_{-0.12}$ and $M_{\rm Walker}/M_{\rm true} =1.07^{+0.21}_{-0.15}$.

%=========================================================
\subsection{Dependence on methods for inferring \Re }
\label{meths}
%=========================================================

In the previous subsection (Figure \ref{r3re}), we showed that the relationship between the inferred $\Re$ of a galaxy and its underlying $r_3$ value is one important source of mass estimator error.  In this subsection we investigate alternative methods for measuring $\Re$ and determine which of these provides the most robust variable to use in the Walker and Wolf estimators.   The fiducial approach we have adopted in proceeding sections -- to measure $\Re$ via a S\'{e}rsic fit to circularly-symmetric bins -- does best, and this is the reason we have used it above. 

\begin{figure}
	\hspace{-0.4cm} \includegraphics[scale=0.22]{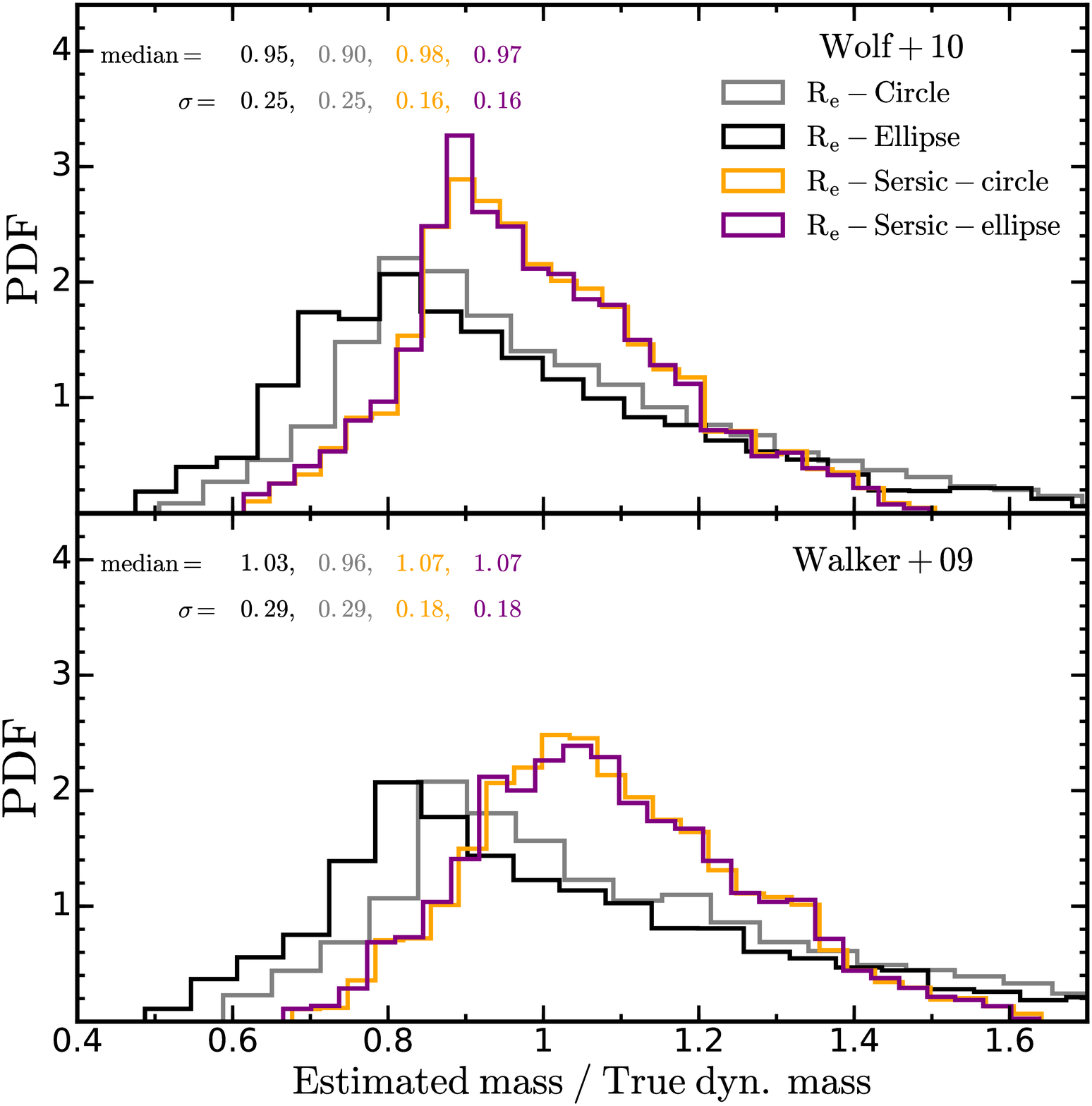}
    \caption{ Same as Figure \ref{stacking} but now the distributions are shown for various methods for determining \Re .  The black and grey histograms use values of \Re\ measured by directly counting the cumulative projected stellar mass profiles and finding the radius where half the total mass is reached.  The colored lines use S\'{e}rsic fits to determine \Re.  The grey/orange lines use circular radii and the black/magenta lines use elliptical radii.  The choice of elliptical vs. circular does not significantly affect results, however fitted profiles are much preferred.}
    \label{methods}
\end{figure}

When attempting to measure $\Re$ one, first needs to decide how to assign a radius to a distribution that is not necessarily circularly symmetric on the sky.  One common choice is to simply  enforce circular symmetry.  Another is to measure the axis ratio on the sky and circularize elliptical bins via  $R \rightarrow R \sqrt{1-\epsilon}$~ \citep[e.g.,][]{Koposov2015}.    Once we have a definition of radius, we must then chose a method for measuring the 2D half-light radius \Re.  Two common choices are 1) to directly count half the light using concentric radial bins or 2) to perform an analytic fit to the projected distribution and to infer \Re\ based on fit parameters.  We have explored all four of these methods to determine \Re\ for our galaxies: both circular and ellipsoidal bins with direct counts and S\'{e}rsic fits.  The results for our stacked samples over all projections are shown in Figure \ref{methods} for each of the four methods.  These results indicate that measurement of \Re\ via an analytic fit is  preferred over direct measurement of the half-light radius.\footnote{For the Wolf estimator, the fitting method does better in finding an \Re\ that obeys $\r3 = 4/3 \Re$.}  The $1\,\sigma$ scatter about the median falls from $0.25$ to $0.16$ ($0.29$ to $0.18$) for the Wolf (Walker) estimator as one moves from direct counts to fitted \Re.  

Figure \ref{methods} also shows that the choice of circular vs. ellipsoidal bins is not very important when $\Re$ values are determined via fit.  The circular approach does slightly better when \Re\ is inferred via direct counts.  This is somewhat surprising given the work of \citet{sanders+2016}, who showed analytically that ``circularized'' half-mass radii \Re$\sqrt{1-\epsilon}$ are preferred for flattened stellar distributions. Particularly, because these authors worked under the assumption that stars and dark matter were stratified along the same self-similar concentric ellipsoids.  As shown in Figure \ref{fig:shape}, our simulated dwarfs seem to obey this assumption on average, though individual systems are not perfectly self-similar, with axis ratios in stars to dark matter $(c/a)_{*}/(c/a)_{DM}$ that range from $\sim 0.8$ to 1.2.

%=========================================================
\section{Conclusions and Discussion}
\label{discuss}
%=========================================================

We have used a suite of twelve state-of-the-art cosmological simulations of dwarf galaxies to test the mass estimators of \citet[][]{Walker+2009} and \citet[][]{Wolf+2010}, which are frequently applied to dispersion-supported dwarf galaxies in the literature.    The simulated dwarfs used in this comparison were presented first in \citet{fitts+2016}.  They have stellar masses $\mstar = 5\times 10^5 \msun$ to $1.5 \times 10^{7} \msun$ and are each resolved with more than one thousand star particles and up to tens of thousands of star particles.  Our dwarfs broadly reproduce the observed properties of low-mass galaxies seen in the local universe, including sizes, velocity dispersions, flat line-of-sight velocity dispersion profiles, and negligible rotation support.  The 3D stellar density profiles for these galaxies is well-represented by a single universal form (Eq. ~\ref{sprofile}, Fig. \ref{densities}), with a core-like distribution $\rho_* \propto 1/r^{0.1}$ at small $r$ a steep fall-off $\rho_* \propto 1/r^{4.2}$ at large $r$. The stellar distributions are aspherical, with an approximately constant short-to-long axis ratios $c/a \sim 0.5$ within twice the half-light radius (Fig. \ref{fig:shape}).

When applied to our simulations in mock projections, both the Walker (Eq.  \ref{eqEs1}) and Wolf (Eq. \ref{eqEs3})  estimators successfully recover the total mass for dispersion supported dwarf galaxies with
medians and 16th-84th percentiles given by $M_{\rm Walker}/M_{\rm true} =1.07^{+0.21}_{-0.15}$ and  $M_{\rm Wolf}/M_{\rm true} = 0.98^{+0.19}_{-0.12}$ when averaged over projection angle for all of our simulated dwarfs.   In a dwarf-by-dwarf comparison, we found that the Walker estimator is biased slightly high ($\sim 10\%$) in the median (Fig. \ref{triaxial}) and has a slightly higher dispersion about the median than the Wolf estimator.  We found that it is important to measure the half-light radius $\Re$ using an analytic fit to the projected density profile rather than cumulative binning in order to achieve an accuracy better than $\sim 25\%$ (Fig. \ref{methods}). The reason is that part of the uncertainty in the mass estimators comes from the mapping between the projected and 3D stellar distributions (Section~\ref{inSims}), and the analytical fits to the projected profiles provide more robust determinations of \Re\ (and \r3).

We  find no correlation between the median accuracy of these mass estimators and the triaxial distribution of stars.  However, for individual galaxies viewed over many projection angles, the dispersion about the median increases as galaxies become more flattened (Fig. \ref{triaxial}).  For one of the most flattened dwarfs ($c/a \approx 0.4$) the $1\,\sigma$ dispersion is $\pm 25\%$ ($\pm  30\%$) for the Wolf (Walker) estimator, while it decreases to $\pm 5\%$ ($\pm 7$) for the most spherical of our galaxies ($c/a \approx 0.7$).  Unfortunately, we find that the observed axis ratio on the sky is not a good indicator of the underlying dispersion about the mean (see Appendix \ref{shape}), so it will be difficult to estimate the expected uncertainty owing to projection by observing the shape on the sky.

Recently,  \citet{Campbell+2016} made use of the APOSTLE simulations of the Local Group to test the same mass estimators explored here against both dispersion-supported galaxies and galaxies with significant rotational support.  They also found that the Walker and Wolf estimators were accurate, especially for their dispersion-supported galaxies.  The dispersion over their entire sample was 25 and 23 percent, respectively. We have found similar qualitative results here, but we find a scatter about the median that is significantly lower: 18 and 16 percent. The lower dispersion in our results could be due to differences in the physical processes included in our respective simulations, as well as our focus on dwarf galaxies with \mstar $\lesssim 10^7$ \msun. However, it may have to do with the methods used to measure \Re. Our higher resolution simulations have allowed us to construct and measure \Re\ values directly by profile fits in each projection.  Campbell et al. determined \Re\ by measuring cumulative half-mass radii. When we use a similar method to the one in \citet{Campbell+2016}, we find larger dispersions: 29 and 25 percent, respectively.   As a final point of comparison we mention that \citet{Campbell+2016} provided a new  mass estimator that did better in their simulations than either the Walker or Wolf estimator.  When we tested their estimator, we found it to be biased low and with larger scatter than either of the two primary estimators we considered. We conclude that the mass estimates presented in Wa09 and Wo10 (1) are accurate and (2) do not have errors that are substantially underestimated.

\citet{sanders+2016} explored mass estimators for dispersion-supported galaxies with \Re\ replaced by the ``circularized'' half-mass radius \Re$\sqrt{1-\epsilon}$.  Using an analytic  approach, they showed that this replacement provided a better mass estimator than the un-circularized approach. To do this, they assumed  that ``stars and dark matter are stratified on the same self-similar concentric ellipsoids."  As shown in Figure \ref{fig:shape}, our simulated dwarfs only roughly conform to this assumption, though in detail the shape of the dark matter and stars are not identical, with ratios $(c/a)_{\rm *}/(c/a)_{\rm DM} \simeq 0.8 - 1.2$.  We tested the estimator using the ``circularized half-mass radius'' and we found that \re\  is not well represented by $4/3\; \Re\sqrt{1-\epsilon}$ and, consequently, the scatter around the median (above 20 percent) is larger than the estimators we tested in this work. \citet{sanders+2016} also pointed out that the uncertainty in the estimate might depend on the ellipticity, which is qualitatively similar to our result that dispersion in mass estimator increases with decreasing axis ratio (see Fig. \ref{triaxial}). 

Our overall conclusion is that the mass estimators proposed by both \citet[][]{Walker+2009} and \citet[][]{Wolf+2010} do a better-than-expected job at reproducing the dynamical masses of simulated galaxies, which have stellar masses of $5\times 10^5 \msun$ -- $1.5 \times 10^{7} \msun$. We caution that errors may be larger for more massive galaxies, which often show more rotational support. It is remarkable that both the Walker and Wolf mass estimators are accurate to better than $\sim 20\%$ considering that our simulated systems break the key assumption of spherical symmetry. Errors at this level are not large enough to  explain the inferred mass discrepancies that are at the origin of the Too-Big-to-Fail problem, and furthermore, the unbiased nature of the estimators makes it unlikely that misestimation of dynamical masses is the explanation of the problem.
Going forward it will be useful to extend this comparison to the question of cusp/core measurement.  Specifically, one of the most interesting uses of these estimators was proposed by \citet{Walker+2011}, who used the existence of multiple populations of stars in individual dwarfs to measure masses at two distinct radii and thus constrain the log-slope of the underlying density profile.  The potential to constrain the density profile slope in very low-mass dwarf galaxies could be a key diagnostic of the CDM paradigm (Bullock \& Boylan-Kolchin 2017, in preparation), and thus testing this method against simulated galaxies with both cusps and cores presents an important next step.  This will be the subject of a forthcoming work.

\section*{Acknowledgements}

A.G-S. was supported by a UC-MEXUS Fellowship. JSB and OE were supported by NSF grant AST-1518291.  JSB was also supported by HST theory programs AR-13921, AR-13888, and AR-14282.001 and program number HST-GO-13343.  These HST programs were provided  by NASA through a grant from the Space Telescope Science Institute, which is operated by the Association of Universities for Research in Astronomy, Incorporated, under NASA contract NAS5-26555.
 MBK and AF acknowledge support from the National Science Foundation (grant AST-1517226). MBK was also partially supported by NASA through grant NNX17AG29G and HST theory grants (programs AR-12836, AR-13888, AR-13896, and AR-14282) awarded by the Space Telescope Science Institute (STScI), which is operated by the Association of Universities for Research in Astronomy (AURA), Inc., under NASA contract NAS5-26555. DK was supported by NSF grant AST-1412153 and the Cottrell Scholar Award from the Research Corporation for Science Advancement. CAFG was supported by NSF through grants AST-1412836 and AST-1517491, and by NASA through grant NNX15AB22G.
 
This work used computational resources of the The University of Texas at Austin and the Texas Advanced Computing Center (TACC; http://www.tacc.utexas. edu), the NASA Advanced Supercomputing (NAS) Division and the NASA Center for Climate Simulation (NCCS) through allocation SMD-16-7760, and the Extreme Science and Engineering Discovery Environment (XSEDE, via allocation TG-AST140080), which is supported by National Science Foundation grant number OCI-1053575.
This research made use of $\textsf {astropy}$ \citep{astropy+2013}.

%%%%%%%%%%%%%%%%%%%% REFERENCES %%%%%%%%%%%%%%%%%%

\bibliographystyle{mnras}

\bibliography{references}

%%%%%%%%%%%%%%%%% APPENDICES %%%%%%%%%%%%%%%%%%%%%

\appendix

\section{Projected Shape}
\label{shape}

\begin{figure}
	\includegraphics[scale=0.27]{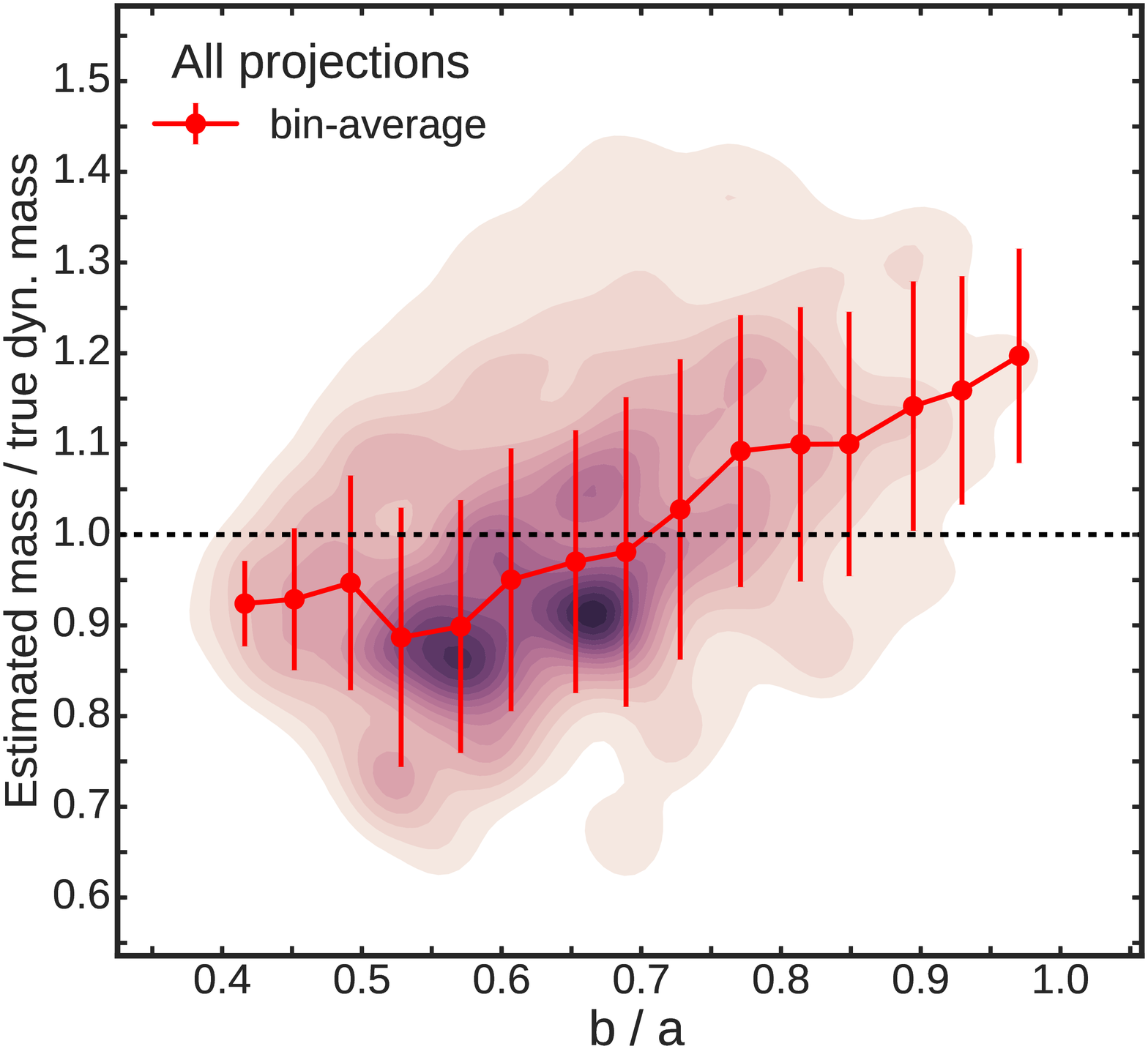}
    \caption{Ratio between the estimated mass when using \citet{Wolf+2010} estimator and the actual dynamical mass within $4/3$\Re\ vs.\  axis-ratio of the projected 2D stellar distribution for all the galaxies. As in Fig. \ref{stacking}, the results for the 1000 projections in all the runs have been stacked (purple smoothed distribution). The distribution has been evenly-binned according to their axis-ratio value, the mean in each bin is showed in red points and the error bars represent their corresponding $1\,\sigma$ scatter. %The solid black line shows the best second-order fit to these points.
 }
    \label{ratios}
\end{figure}

We compute the axis ratios\footnote{Note that that the projected axis values ($a$,$b$) are different from the ones ($a$,$b$,$c$) used in the main text, since the last are calculated by analyzing the 3D distribution.} ($b/a$) for each of the projected distributions in each simulated galaxy. As we found that both estimators gave ``similar'' (see Figure~\ref{stacking}) results, in the following we will focus our analysis in the Wo10 estimator. We use an iteratively method \citep{dubinski+1991} for all the stellar particles within $4/3\Re$ (Wo10). In Figure~\ref{ratios} we plot the ratio between the estimated mass by Wo10 estimator and the actual dynamical mass within $4/3$\Re\ against the axis-ratio of the projected 2D stellar distribution for all the galaxies. We stacked the results for all the dwarf galaxies in our suite of simulations (purple shaded area). In order to have a more clear understanding of how the axis-ratios correlate with the mass estimator, we have evenly-binned the whole distribution according to their axis-ratio value, the mean in each bin is showed in red points and the error bars represent the $1\,\sigma$ scatter. It is clear that for 2D-ellipse axis-ratios larger than 0.7 the Wo10 estimator tend to overestimate the true dynamical mass within $r=4/3\Re$, and the larger the $b/a$ values, the more inaccurate become the estimator, reaching a 25 percent offset in the limit where $b/a\approx 1$. For $b/a$ values smaller than 0.65, the opposite behavior occurs, but in this case Wo10 result in a $\approx$ 10 percent constant underestimation of the true total mass.

We would be tempted to think that $b/a$ values closer to one, reflecting a projected spherical distribution would give better results, taking into account that mass estimators assume this kind of symmetry. Certainly that would be the case if the stellar particles in our simulated galaxies were distributed spherically. However, we found that estimators with $b/a$ values larger than 0.7 almost always overestimate the true dynamical mass in the dwarfs; what this is telling us is that the intrinsic 3D-distribution of the stellar particles is triaxial, and l.o.s. in which $b/a$ is close to one, represent projections that are far form the real stellar distribution. Indeed, the lowest 3D-ellipsoid semi-axes ratios, $c/a$, that gives an idea of how spherical is the 3D-distribution, have values around 0.7 for most of our runs.

%%%%%%%%%%%%%%%%%%%%%%%%%%%%%%%%%%%%%%%%%%%%%%%%%%

% Don't change these lines
\bsp	% typesetting comment
\label{lastpage}
\end{document}